\documentclass[10pt, conference]{IEEEtran}

\usepackage{paralist}
\usepackage{mathptmx}
\usepackage{graphicx}
\usepackage{fancyhdr}
\newcommand{\ProportionOfText}{0.45}
\newcommand{\Tomahawk}{Tomahawk }

\title{GMine: A System for Scalable, Interactive Graph Visualization and Mining}

  \author{%
    \IEEEauthorblockN{Jose F. Rodrigues Jr.*, Hanghang Tong+, Agma J. M. Traina*, Christos Faloutsos+, Jure Leskovec+}
    \IEEEauthorblockA{%
*University of Sao Paulo, Brazil\\
\{junio,agma\}@icmc.usp.br}
+Carnegie Mellon University, USA\\
+\{htong,christos,jure\}@cs.cmu.edu}

\begin{document}
\maketitle

\begin{abstract}
Several graph visualization tools exist. However, they are not able to handle large graphs, and/or they do not allow interaction. We are interested on large graphs, with hundreds of  thousands of nodes. Such graphs bring two challenges: the first one is that any straightforward interactive manipulation will be prohibitively slow. The second one is sensory overload: even if we could plot and replot the graph quickly, the user would be overwhelmed with the vast volume of information because the screen would be too cluttered as nodes and edges overlap each other. GMine system addresses both these issues, by using summarization and multi-resolution. GMine offers multi-resolution graph exploration by partitioning a given graph into a hierarchy of com-munities-within-communities and storing it into a novel R-tree-like structure which we name {\it G-Tree}. GMine offers summarization by implementing an innovative subgraph extraction algorithm and then visualizing its output.
\end{abstract}

\vspace{-0.35cm}\section{Introduction}
\label{Introduction}

An important support for graph exploration is interactive visualization, which can help to quickly identify the main components of a graph, its outliers, the most important edges and communities of related nodes. Interaction-enabled visualization allows to pick detailed and contextualized information on demand, interact with nodes and edges and determine topology aware arrangements for clearer inspection.

However, up-to-date applications have produced graphs on the order of hundred thousand nodes and possibly million edges (referenced from here on as large graphs). Large graphs can be found in numerous real-life settings: web graphs (web pages, pointing to others with hypertext links)~\cite{kumar99trawling}, computer communication graphs (IP addresses sending packets to other IP addresses), recommendation systems~\cite{leskovec06amazon}, who-trusts-whom networks~\cite{richardson03trust}, bipartite graphs of web-logs of who visits what page; blogs and similar. At this magnitude, efficient graph visualization becomes prohibitive because of the excessive processing power requirements that prevent interaction. Besides that, hundred-thousand-node drawings result in unintelligible cluttered images that do not aid the user’s cognition. To face these challenges we present a system that explores two new ideas to address scalability in large graph visualization. The first idea establishes a hierarchical partitioned arrangement from a graph in order to allow multi-resolution visualization. The second idea utilizes an innovative algorithm to extract a subgraph of interest based on an initial set of target nodes. Our system uses either or both of these ideas to process large graphs bypassing the aforementioned limitations of massive graph drawing. The proposed interface permits to navigate through the levels of a graph hierarchy and also to mine subgraphs information for targeted graph exploration.

The remaining of this paper is structured as follows. Section \ref{Dataset} introduces the DBLP dataset that will be used along this work. Section \ref{Hierarchical} describes our multi-resolution visualization idea and section \ref{SubGraph} illustrates our subgraph extraction algorithm. Section \ref{Conclusions} concludes the work.

\vspace{-0.35cm}\section{DBLP dataset}
\label{Dataset}

Throughout this text we employ the DBLP dataset to illustrate the functionalities of our system. This dataset originates from the Digital Bibliography \& Library Project (or DBLP). DBLP is a publicly available database of publication data that embraces authors (also co-authors) from the Computer Science community and their published works. Its content is periodically updated and detailed information from DBLP can be achieved at {\it http://dblp.uni-trier.de/}.
The version of DBLP dataset that we use defines a graph with $n=315,688$ nodes and $e=1,659,853$ edges, where each node represents an author of a publication and each edge denotes a co-authoring relationship between two authors.
      
\vspace{-0.35cm}\section{Graph Hierarchy Creation,\\
Structuring and Visualization}
\label{Hierarchical}

Our first idea to deal with massive graphs is the use of a commu-nities-within-communities structured visualization.  In the next sections we overview the steps to come up with such proposal at the same time that we describe its features for visualization and interaction.

\subsection{The G-Tree structure}

For this work, initially we need to recursively and hierarchically partition a given graph. We adopted the methodology named {\it k-way} partitioning (however any partitioning methodology fits our system). That is, given a graph $G = (V, E)$ with $|V| = n$, we want to have $k$ subsets $V_1, V_2, . . ., V_k$ of $V$, such that $V_i \cap V_j = \emptyset$ for $i \neq j$, $|V_i| = n/k$ and $\cup _i V_i = V$. Also, the partitioning must minimize the number of edges of $E$ whose incident vertices belong to different subsets. This partitioning methodology is implemented by METIS, whose details are found in the work by Karypis and Kumar \cite{1} and in related works.

\begin{figure}[htb]
	\centering	   
\includegraphics[width=0.47\textwidth]{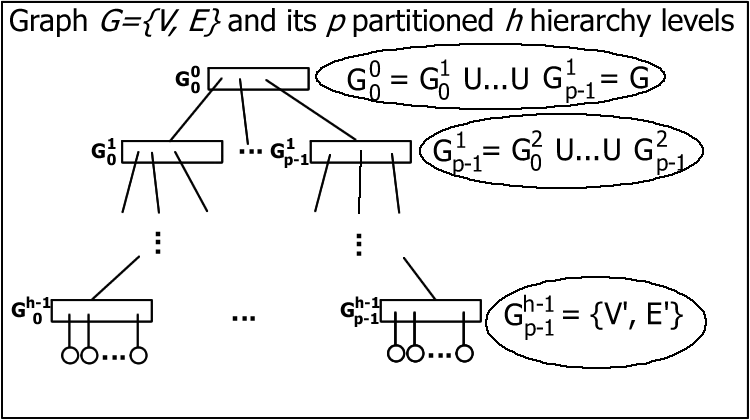}
	\caption{G-Tree structure, which we utilize for our visualization system together with its graph recursive structuring.}
	\label{R-Tree}
\end{figure}

Hence, given a graph, we perform a sequence of recursive partitionings to achieve a hierarchy of communities-within-communities. At each recursion, each partition is submitted to a new partitioning cycle that will create another set of partitions. This process repeats until we get the desired granularity for the partitions (communities). For each new set of partitions, a new subtree is embedded in an R-tree like structure. At each new level of the tree, the tree nodes (communities) just created will have the formerly partitioned tree node as their parent. We call this structure G-Tree (named after Graph-Tree), which is the data structure that supports our system, illustrated in figure \ref{R-Tree}. The references for the graph nodes properly said are at the bottom level of the tree. The entire structure is stored in a single file and the nodes are transferred to main memory only when necessary.

To demonstrate our methodology, we recursively partition DBLP dataset into $5$ hierarchy levels each with $5$ partitions. The dataset, thus, is broken into $5^4 + 1$, or $626$, communities with an average of $500$ nodes per community. The communities reflect the connectivity (number of edges) among their members according to METIS partitioning algorithm.

\subsection{Visualization and Interaction}

We propose an innovative interactive presentation for large graphs. For this purpose, our system promotes the navigation across the levels of the tree that represents the hierarchical partitioning of a large graph. As the user interacts with the visualization, the system keeps track of the connectivity among communities of nodes at different levels of the partitioned graph. When the user changes the focus position on the tree structure, the system works on demand to calculate and present contextual information.

\begin{figure}[htb]
	\centering	   
\includegraphics[width=0.47\textwidth]{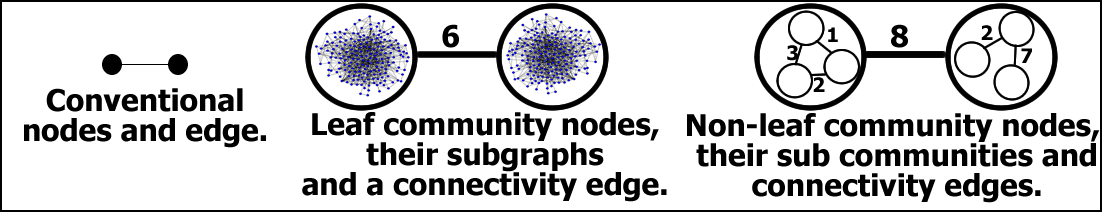}
	\caption{Conventional nodes and an edge to denote relationship. Leaf community nodes, subgraphs and a connectivity edge to denote how many nodes from the communities have an edge to connect them. Non-leaf community nodes, sub communities and connectivity edges.}
	\label{NodesAndEdges}
\end{figure}

When we display a graph as communities-within-communities, we have new representations for graph drawing, as illustrated in figure \ref{NodesAndEdges}. Besides conventional nodes and edges that appear only at the bottom level of the tree (leaf nodes), we also have community nodes, that comprehend a number of sub communities and nodes, and we have connectivity edges, that represent the number of edges between community nodes. These connectivity edges represent the number of edges between nodes from the original graph, but that are in different communities. The storage and management of this information is out of the scope of this demonstration paper.

\begin{figure}[htb]
	\centering	   
\includegraphics[width=\ProportionOfText\textwidth]{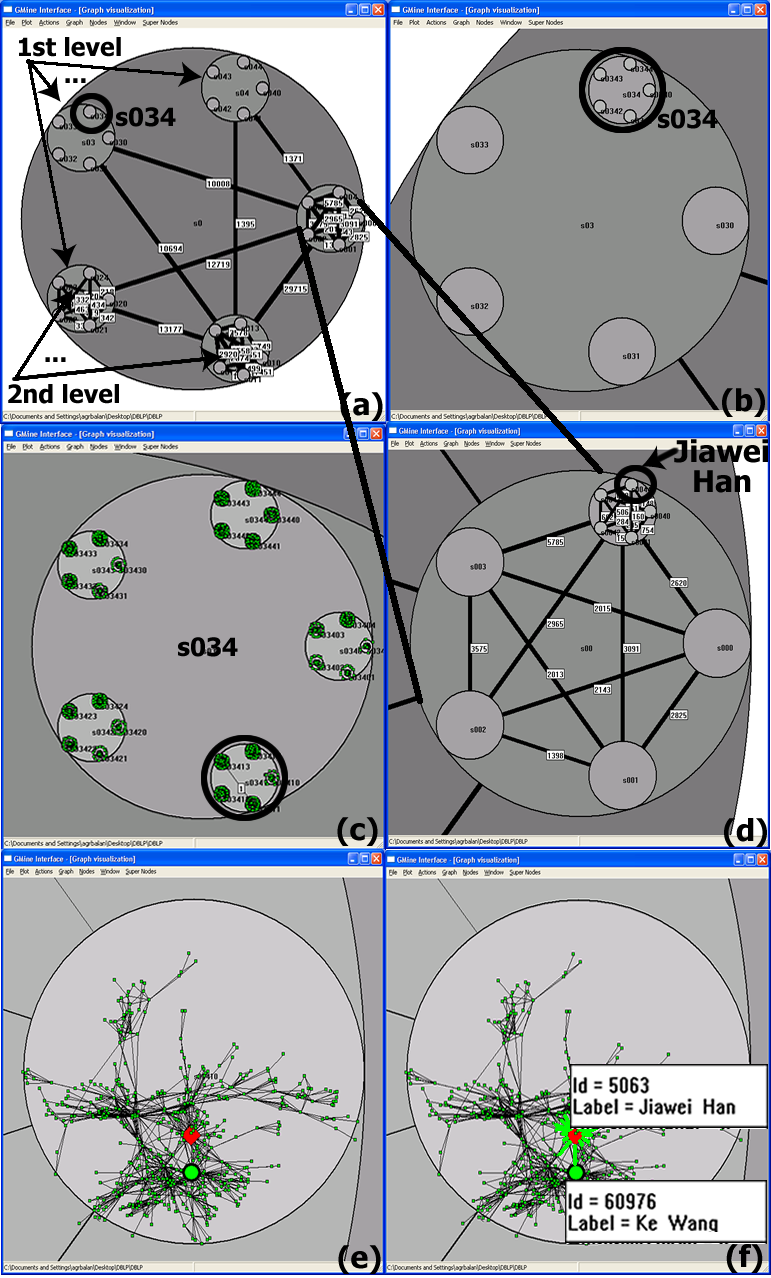}
	\caption{DBLP dataset navigation. (a) First $5$ main communities and its $25$ sub communities. (b) Contextualization of community {\it s034}. (c) Closer look and complete expansion of community {\it s034}. (d) We locate author Jiawei Han. (e) Subgraph community of author Jiawei Han. (f) Interaction with the subgraph reveals co-author Ke Wang as one of the main contributors to Jiawei Han.}
	\label{DBLPComplete}
\end{figure}

These features are illustrated in figure \ref{DBLPComplete}, which presents a sequence of interactive actions taken by the user when navigating in DBLP dataset. In figure \ref{DBLPComplete}(a), it is possible to see DBLP partitioned into $5$ communities in its first hierarchy level, and other $5*5$, or $25$ communities in its second hierarchy level. At this point, $3$ communities are highly connected to every other community and also highly connected among their $5$ sub communities. The other $2$ first-level communities are relatively isolated from the other $3$ and totally isolated among their sub communities. One can conclude that the $3$ highly connected communities hold long term active and collaborating authors, while the other $2$ hold casual, less productive authors who seldom interact with each other. In figure \ref{DBLPComplete}(b) we focus on community {\it s034} and verify that its sub communities are isolated from each other. A deeper focus in community {\it s034} in figure \ref{DBLPComplete}(c) shows that among its sub communities (highlighted), only two of these sub communities present an edge. Our system allows to inspect this specific outlier edge to reveal that authors ``D. B. Miller'' and ``R. G. Stockton''  define this co-authoring relation for their unique DBLP publication dated from $1989$. It is also possible to execute a label query to locate a specific author within the hierarchy, as for example author Jiawei Han in figure \ref{DBLPComplete}(d). In figure \ref{DBLPComplete}(e) we go to its subgraph community and verify other important nodes surrounding this author. In figure \ref{DBLPComplete}(f) we interact with the graph to discover author Ke Wang, which is another very active author who has worked for years with author Jiawei Han.

The exploration of communities of nodes instead of all the nodes at a time, the way we are doing, allows the perception of the relationships among communities of nodes. This way it is possible to trace the distribution of edges among communities, their connectivity degree and their scope of connectivity. It is also possible to pick outlier edges for suspicious connections between communities. The user can focus at different communities of nodes according to his/her interest and browse the levels of the hierarchy in order to identify interesting connections or to inspect specific graph nodes.

At the bottom level of the tree, the user can access a subgraph that is part of the larger graph being analyzed. To do so, the system brings the correspondent graph nodes from disk to memory and draws them inside the region attributed to its parent community (tree node). Then this area of the visualization scene becomes a regular area for graph drawing. For this subgraph, besides basic interaction (zoom, pan and details on demand) the user can also ask for the calculation of metrical features corresponding to this subgraph only. Our system supports the following calculations: degree distribution, number of hops, number of weak components, number of strong components and page rank calculation for the nodes. GMine also offers pop up node information, edge expansion and edition of nodes and edges.

\subsection{The \Tomahawk Principle}

\begin{figure}[htb]
	\centering	   
\includegraphics[width=0.44\textwidth]{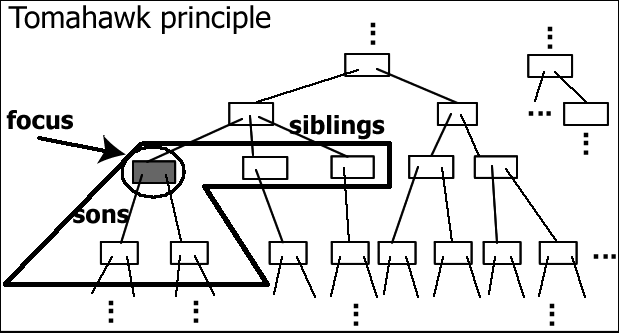}
	\caption{The \Tomahawk principle to help decide what to exhibit according to user interaction.}
	\label{Tomahawk}
\end{figure}

The presentation of the node communities together with the edges that connect them may cause sensory overload. This is due to the fact that every community can potentially be connected to every other community. This problem is aggravated if the graph has many hierarchy levels exhibited simultaneously when communities are expanded to show their content.
To cope with this aspect of our multi-resolution graph visualization, we propose to display a small, but carefully chosen set of communities. We refer to this method as the ``\Tomahawk'' principle, because the chosen nodes remind of a tomahawk ax when shown on G-Tree method, illustrated in figure \ref{Tomahawk}. That is, in order to limit the number of items presented at a time, we make use of G-Tree structure to determine a well-established context every time in response to user interaction. Thus, as the user chooses a community node to focus on, we traverse the tree in order to gather the desired node of interest, its sons and its siblings. Then we plot only these items inside the minimum node that bears this contextualization, see figure \ref{DBLPComplete}(b). We argue that the \Tomahawk principle can provide a minimum contextualization to the user by presenting nodes above, beneath and by the side of a node of interest. 

\vspace{-0.35cm}\section{Connection Subgraph \\
Extraction}
\label{SubGraph}

Our second idea to deal with massive graphs is the use of a novel algorithm for connection subgraph extraction. Our algorithm, which is not to be detailed in this demonstration work,
aims to maximize what we call ``goodness score'' of the nodes within a subgraph. To reach this goal, an independent random walk with restart is simulated for each source node, and the goodness score of a node is computed by the steady-meeting probability that the random particles will finally meet each other at the given node. Then, a dynamic programming is used to discover important paths iteratively. The proposed algorithm can deal with multi-source queries, while the existing one ~\cite{Christos04conngraph} is restricted to pairwise source queries.

A typical scenario to apply connection subgraph extraction is ``given an initial set of interesting individuals, find a small number of individuals from a large social network that can best capture the relationship among the individuals of the initial set''. For large graphs, extracting a small (say, with tens of nodes) yet representative connection subgraph brings feasibility to large graph visual exploration. Also, due to the multi-faced nature of many real life relationships, connection subgraphs provide a better way to describe such kind of relationships if compared to single path descriptions.

\begin{figure}[htb]
    \centering
\includegraphics[width=0.46\textwidth]{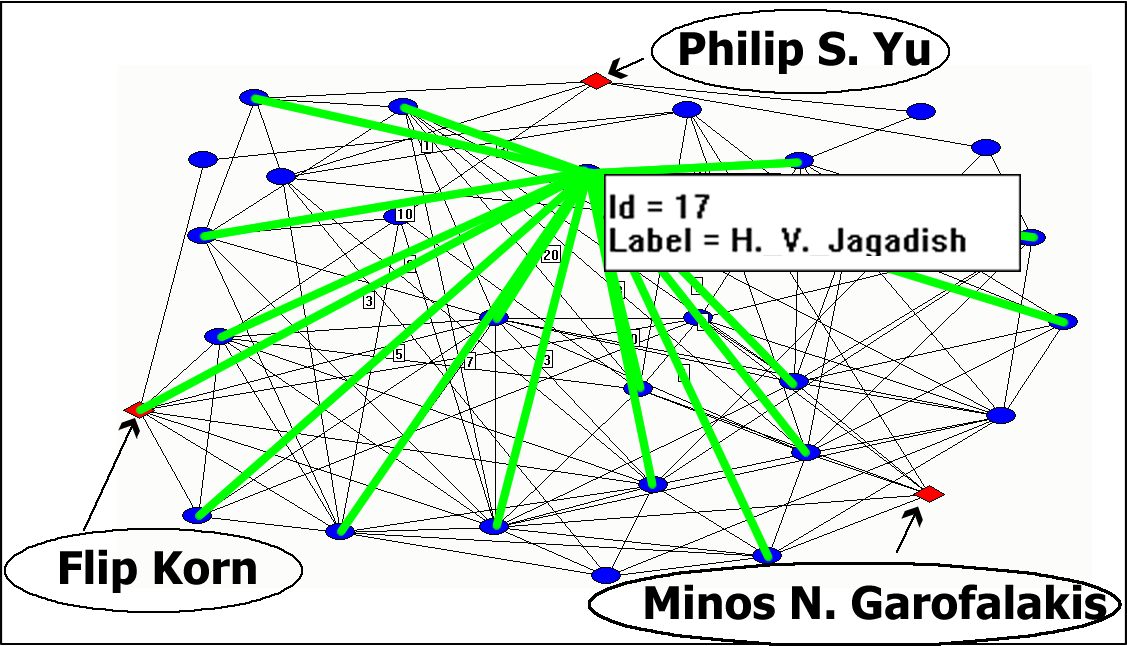}
    \caption{Illustration for connection subgraph extraction.}
    \label{connsubgraph}
\end{figure}

For (limited static) demonstration, a connection subgraph with $30$ nodes extracted from the whole
DBLP dataset is plotted in figure~\ref{connsubgraph}. The initial query set in figure~\ref{connsubgraph} is composed of three authors from the database community: ``Philip S. Yu'', ``Flip Korn'' and ``Minos N. Garofalakis''. In figure~\ref{connsubgraph}, instead of a thousand nodes graph,
one can concentrate on a subgraph of interest extracted from the original graph. The magnitude of the subgraph is thousand fold smaller than the original dataset and the subgraph being visualized is directly related to the interconnection defined by our initial set of target nodes.

On the visualization, if the user moves the mouse over a node, GMine pops up more information about that node -
in the example, one can see Prof. H. V. Jagadish data and his edges highlighted. Prof. Jagadish has direct connection with Flip Korn, and 1-step-away connections with Dr. Philip Yu and Dr. Minos Garofalakis.

\begin{figure}[htb]
	\centering	   
\includegraphics[width=0.46\textwidth]{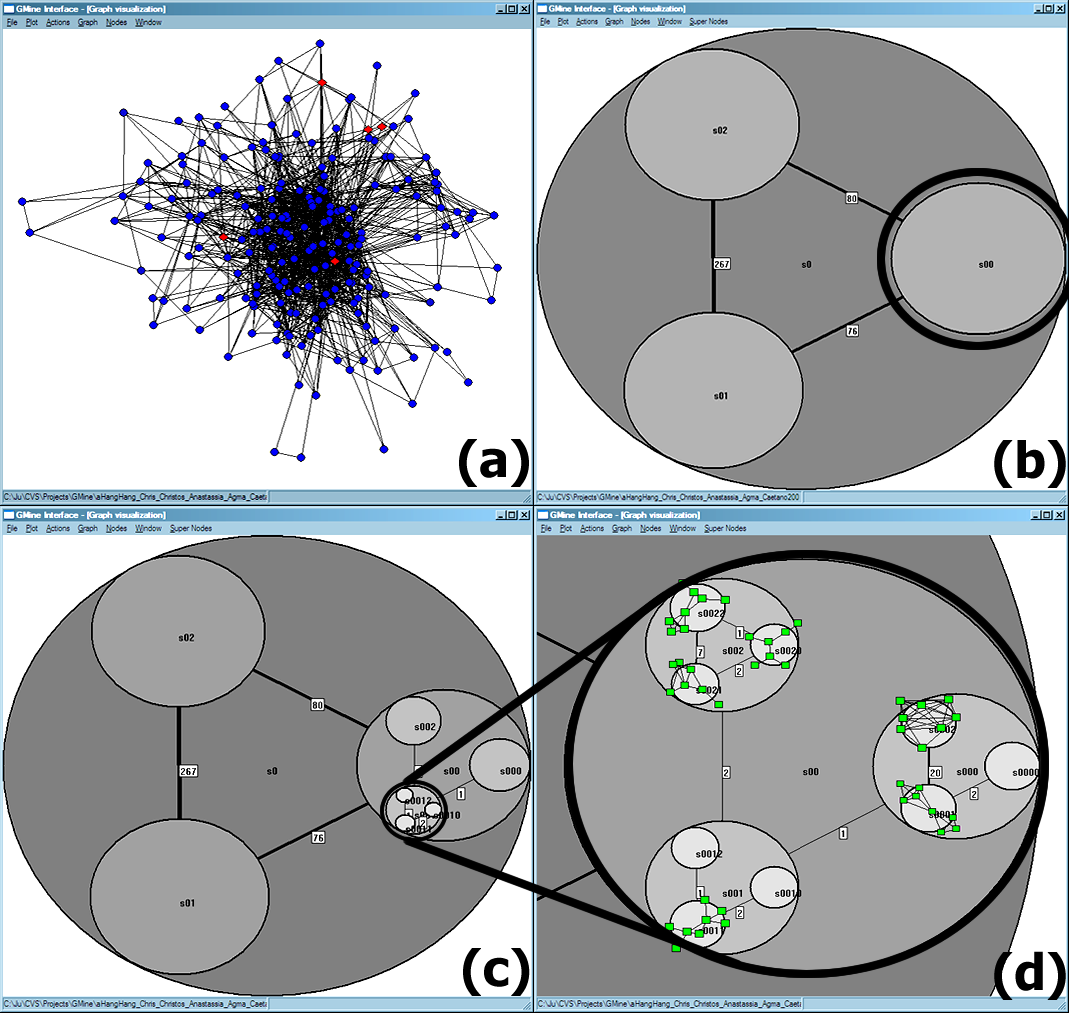}
	\caption{(a) A $200$ nodes subgraph extracted from DBLP dataset. (b) The same graph presented as three partitions. (c) One level down the hierarchy and we have three other communities inside the community highlighted in (b). (d) Zoom in the community highlighted in (c) and another level down the hierarchy. We reach the very nodes of the graph.}	\label{DBLPPartial}
\end{figure}

In our system, subgraph extraction can be utilized alone or combined to communities-within-communities visualization. Alone, one can extract a subgraph of interest from a given large graph. Combined, (see figure \ref{DBLPPartial}), it can be used to generate a subgraph to be hierarchically partitioned for visualization or, alternatively, it can be used to generate a subgraph from an existing graph partition.

Figure \ref{DBLPPartial} illustrates the combination of subgraph extraction and com-munities-within-communities visualization. Figure, \ref{DBLPPartial}(a) displays a $200$ nodes subgraph extracted from the DBLP dataset. In figure \ref{DBLPPartial}(b) it possible to see this subgraph partitioned into $3$ main communities. In figures \ref{DBLPPartial}(c) and \ref{DBLPPartial}(d) we go deeper into the hierarchy to analyze the connectivity between communities and, finally, the very nodes of the graph.

\vspace{-0.35cm}\section{Conclusions}
\label{Conclusions}

We have demonstrated a system that supports the visualization of large graphs in an interactive environment. In our tool the user can navigate through the graph structure in a hierarchical fashion, having different perspectives of the graph arrangement, varying from multiple resolution levels to detailed inspection of specific graph nodes. The system also supports an innovative subgraph extraction algorithm that can speed up large graph exploration by concentrating on a targeted subset of the graph.

The benefits of our ideas come from its compartmented graph management that promotes scalability while keeping visual comprehension. The scalability is due to the fact that smaller parts of the graph are processed one at a time instead of the whole graph at every cycle. Visual comprehension derives from limited visual data presentation in contrast to cluttered visualizations generated when large graphs are entirely drawn.

Due to space limitations it is not possible to show all the GMine functionalities. Therefore, for a better exposition, we have GMine available online at \ \ {\it \verb+http://www.cs.cmu.edu/~junio/GMine+}, where the software, datasets and videos can be downloaded. For	 VLDB demonstration session, we plan to let the interested VLDB participants interact directly with the system, possibly checking for their name, their connection-subgraphs with their colleagues, and zooming in and out their corresponding communities.

\vspace{-0.35cm}\section{Acknowledgements}
\noindent{This work has been supported by FAPESP (S\~ao Paulo State Research Foundation), CNPq (Brazilian National Research Foundation), CAPES (Brazilian Committee for Graduate Studies), National Science Foundation, (PITA) Pennsylvania Infrastructure Technology Alliance and donations from Intel, NTT and Hewlett-Packard. Any opinions, findings and conclusions or recommendations expressed here are those of the author(s) and do not necessarily reflect the views of the funding parties.}

\vspace{-0.3cm}    
\bibliographystyle{plain}

\end{document}